\documentclass[
11pt,%
tightenlines,%
twoside,%
onecolumn,%
nofloats,%
nobibnotes,%
nofootinbib,%
superscriptaddress,%
noshowpacs,%
centertags]%
{revtex4}
\usepackage{ljm}
\begin{document}

\titlerunning{algorithm for solving linear inequalities} % for running heads
\authorrunning{L. B. Sokolinsky, I. M. Sokolinskaya} % for running heads

\title{Scalable Parallel Algorithm for Solving
Non-stationary Systems \\ of Linear Inequalities}

\author{\firstname{L.~B.}~\surname{Sokolinsky}}
\email[E-mail: ]{leonid.sokolinsky@susu.ru}
\affiliation{South Ural State University (National Research University), Lenin prospekt, 76, Chelyabinsk, 454080 Russia}

\author{\firstname{I.~M.}~\surname{Sokolinskaya}}
\email[E-mail: ]{irina.sokolinskaya@susu.ru}
\affiliation{South Ural State University (National Research University), Lenin prospekt, 76, Chelyabinsk, 454080 Russia}

\firstcollaboration{(Submitted by E.~E.~Tyrtyshnikov)} % Add if you know submitter.

\received{March 10, 2020; revised March 23, 2020; accepted March 31,
2020}

 % The date of receipt to the editor, i.e. December 06, 2017

\begin{abstract} % You shouldn't use formulas and citations in the abstract.
In this paper, a scalable iterative projection-type algorithm for
solving non-stationary  systems of linear inequalities is
considered. A non-stationary system is understood as a large-scale
system of inequalities in which coefficients and constant terms can
change during the calculation process. The proposed parallel
algorithm uses the concept of pseudo-projection which generalizes
the notion of orthogonal projection. The parallel pseudo-projection
algorithm is implemented using the parallel BSF-skeleton. An
analytical estimation of the algorithm scalability boundary is
obtained on the base of the BSF cost metric. The large-scale
computational experiments were performed on a cluster computing
system. The obtained results confirm the efficiency of the proposed
approach.
\end{abstract}

 % Enter 2010 Mathematics Subject Classification.

 % Include keywords separeted by comma.

\maketitle

% Text of article starts here.

\section{Introduction}

The problem of finding a feasible solution to a linear inequality system also known as a feasibility problem is often encountered in the practice of mathematical modeling. As examples, we can mention linear programming \cite{1,2}, image reconstruction from projections \cite{3}, tomography image reconstruction~\cite{4} and intensity modulated radiation therapy \cite{5}. In many cases, the linear inequality systems arising in such context involve up to tens of millions of inequalities and up to hundreds of millions of variables \cite{2}. Moreover, in mathematical economic models, systems of linear inequalities are often non-stationary. It means that the coefficients of the system and constant terms are changed during the process of solving the problem, and the period of changing the source data can be within hundredths of a second.

At present time, there are a lot of methods  for solving systems of
linear inequalities. Among these methods, we can distinguish a class
of ``self-correcting'' iterative methods that can be parallelized
efficiently. Pioneering works here are papers \cite{6,7}, which
propose the Agmon--Motzkin--Schoenberg relaxation method for solving
systems of linear inequalities. This method uses the orthogonal
projection onto a hyperplane in Euclidean space. Censor and Elfving
in \cite{3,8} proposed a modification of the Cimmino method
\cite{9,10,11} for solving systems of linear inequalities in
Euclidean space $\mathbb{R}^n$. The similar method of
pseudo-projections based on Fejer approximations was proposed by the
authors in \cite{12}. In the article \cite{2}, the pseudo-projection
method was used to solve the problem of finding a feasible solution
to a non-stationary linear inequality system. The convergence
theorem for this method was proven by the authors for the case when
changing the feasible set is a translation. We have constructed a
parallel implementation of the pseudo-projection method and executed
large-scale computational experiments on a cluster computing system
by varying the displacement rate of the polytope~$M$ bounding the
feasible region. Performed evaluation showed that the parallel
pseudo-projection algorithm converges only with very low rate of
displacement of the polytope $M$.

The aims of this article are the following: analyzing the low
efficiency  of the parallel pseudo-projection algorithm for the
non-stationary case, modifying the algorithm to solve this issue,
evaluating the modified algorithm and conducting the large-scale
computational experiments on a cluster computing system to examine
the efficiency of proposed solution.

The paper is organized as follows. In Section \ref{section2}, we
provide a  formal definition of a non-stationary system of linear
inequalities and describe a modified pseudo-projection algorithm
ModAP calculating a feasible solution for such systems under
condition of source data dynamic changes. Section \ref{section3}
describes the ModAPL algorithm, which is a representation of the
ModAP algorithm in the form of operations on lists using the
higher-order functions $Map$ and $Reduce$, and presents a parallel
implementation of the ModAPL algorithm. Section \ref{section4} is
devoted to an analytical evaluation of the ModAPL parallel algorithm
scalability by using the cost metric of the BSF parallel computation
model. Section \ref{section5} provides an information about the
implementation of the ModAPL parallel algorithm, as well as
describes the results of large-scale computational experiments on a
cluster computing system that confirm the efficiency of the proposed
approach. Section \ref{section6} summarizes the obtained results and
concludes that the scalability of the algorithm depends on the
number of dynamically changing parameters of the source linear
inequality system.

\section{Non-stationary problem and pseudo-projection \protect\\ algorithm} \label{section2}

Let the following feasible system of linear inequalities be given in ${\mathbb{R}^n}$:
\begin{equation}\label{Formula1}
{A^{(t)}}x \leq {b^{(t)}},
\end{equation}
where the matrix ${A^{(t)}}$ has $m$ rows. The
non-stationarity of the system~\eqref{Formula1} is understood in the
sense that the entries of the matrix ${A^{(t)}}$ and the elements of
column ${b^{(t)}}$ depend on time $t \in {\mathbb{R}_{ \geq 0}}$.
Let ${M^{(t)}}$ be a polytope bounding the feasible region of the
system~\eqref{Formula1} at instant of time $t$. Such a polytope is
always a closed convex set. We will also assume that the polytope
${M^{(t)}}$ is a bounded set. Let us define the distance from the
point $x \in {\mathbb{R}^n}$ to the polytope ${M^{(t)}}$ as follows:
\begin{equation*}\label{Formula2}
d\left( {x,{M^{(t)}}} \right) = \mathop {\inf }\limits_{y \in {M^{(t)}}} \left\| {x - y} \right\|,
\end{equation*}
where $\left\|  \cdot  \right\|$ signifies the Euclidean norm. Let
us denote the $i$-th row of the matrix $A^{(t)}$ as $a_i^{(t)}$. We assume from now on that $a_i^{(t)}$ is not equal to the
zero vector for all $i = 1, \ldots ,m$. Let $P_i^{(t)}$ be a half-space
representing a set of feasible points for the $i$-th inequality of
the system~\eqref{Formula1}:
\begin{equation*}\label{Formula3}P_i^{(t)} = \left\{ {x|x \in {\mathbb{R}^n},\left\langle {a_i^{(t)},x} \right\rangle  \leq b_i^{(t)}} \right\}.
\end{equation*}
Then, ${M^{(t)}} = \bigcap\limits_{i = 1}^m {P_i^{(t)}}.$ Each
equation $\left\langle {a_i^{(t)},x} \right\rangle  = b_i^{(t)}$
defines the corresponding hyperplane $H_i^{(t)}$:
\begin{equation*}\label{Formula4}
H_i^{(t)} = \left\{ {x \in {\mathbb{R}^n}|\left\langle {a_i^{(t)},x} \right\rangle  = b_i^{(t)}} \right\}.
\end{equation*}
We define the \emph{reflection vector} ${\rho _{H_i^{(t)}}}(x)$ of the point $x$ with respect to the hyperplane ${H_i}$ as follows:
\begin{equation*}\label{Formula5}
{\rho _{H_i^{(t)}}}(x) = \frac{{\left\langle {a_i^{(t)},x} \right\rangle  - b_i^{(t)}}}{{{{\left\| {a_i^{(t)}} \right\|}^2}}}a_i^{(t)}.
\end{equation*}
Then, the \emph{orthogonal projection} ${\pi _{H_i^{(t)}}}(x)$ of
the point $x$ onto hyperplane $H_i^{(t)}$ is calculated by the
%following
equation
%\begin{equation}\label{Formula6}
${\pi _{H_i^{(t)}}}(x) = x - {\rho _{H_i^{(t)}}}(x).$%\end{equation}

\begin{figure}
    \setcaptionmargin{40mm}
    \onelinecaptionsfalse % if the caption is multiline
    %\onelinecaptionstrue  % if the caption is one-line
    \centering
    \includegraphics[scale=0.8]{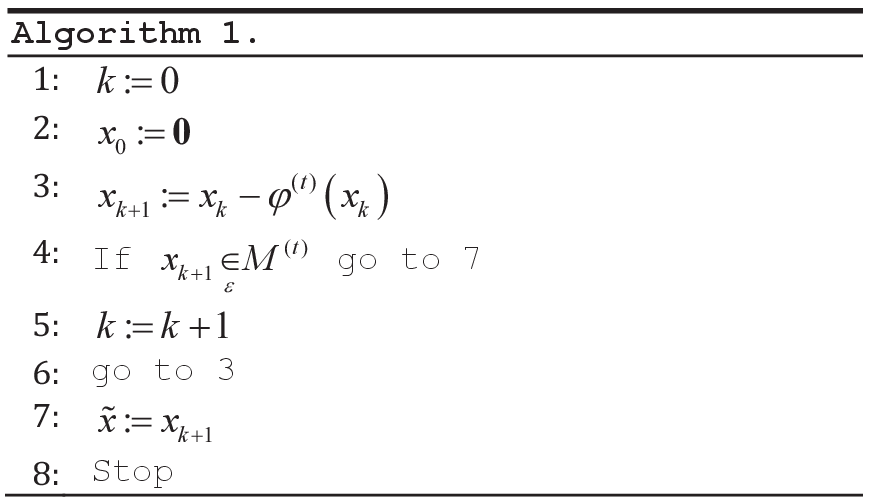}
    \captionstyle{normal}\caption{Pseudo-projection algorithm for non-stationary systems of linear inequalities.}\label{Fig1}
\end{figure}

Define the vector $\rho _{H_i^{(t)}}^ + (x)$ as a \emph{positive
slice} of the reflection vector
\begin{equation}\label{Formula7}\rho _{H_i^{(t)}}^ + (x) = \frac{{\max \left\{ {\left\langle {a_i^{(t)},x} \right\rangle  - b_i^{(t)},0} \right\}}}{{{{\left\| {a_i^{(t)}} \right\|}^2}}}a_i^{(t)}.\end{equation}
The vector $\rho _{H_i^{(t)}}^ + (x)$ will be a non-zero vector if and only if the point $x$ does not satisfy the $i$-th inequality of the system~\eqref{Formula1}.
Let us designate
\begin{equation}\label{Formula8}{\varphi ^{(t)}}\left( x \right) = \frac{1}{h}\sum\limits_{i = 1}^m {\rho _{H_i^{(t)}}^ + (x)} ,\end{equation}
where $h$ is the number of non-zero terms in the sum $\sum\limits_{i = 1}^m {\rho _{H_i^{(t)}}^ + (x)} $.

Define the half-space ${P_i}$ $(i = 1, \ldots ,m)$ as follows:
%\begin{equation}\label{Formula9}
$P_i^{(t)} = \left\{ {x \in {\mathbb{R}^n}|\left\langle
{a_i^{(t)},x} \right\rangle  \leq b_i^{(t)}}
\right\}.$
%\end{equation}
We will say that the point $\tilde x$
belongs to the half-space $P_i^{(t)}$ with precision $\varepsilon  >
0$ and denote it as $\tilde x\mathop  \in \limits_\varepsilon
P_i^{(t)}$, if the following condition holds
\begin{equation}\label{Formula10}\forall i \in \{ 1, \ldots ,m\} \left( {\tilde x \in P_i^{(t)} \vee \frac{{\left|
{\left\langle {a_i^{(t)},\tilde x} \right\rangle  - b_i^{(t)}} \right|}}{{\left\| {a_i^{(t)}} \right\|}} < \varepsilon } \right).
\end{equation}
This means that, for any $i = 1, \ldots ,m$, either the point
$\tilde x$ belongs to the half-space $P_i^{(t)}$, or  the distance
between the point $\tilde x$ and this half-space is less than
$\varepsilon $. We assume that the point $\tilde x$ belongs to the
polytope ${M^{(t)}}$ with precision $\varepsilon $, and we will
denote it as $\tilde x\mathop  \in \limits_\varepsilon  {M^{(t)}}$,
if the following condition holds:
\begin{equation*}\label{Formula11}
\forall i \in \{ 1, \ldots ,m\} \left( {\tilde x\mathop  \in
\limits_\varepsilon  P_i^{(t)}} \right).
\end{equation*} In other
words, for any $i = 1, \ldots ,m$, either the point $\tilde x$
belongs to the polytope ${M^{(t)}}$, or the distance between the
point $\tilde x$ and the polytope ${M^{(t)}}$ is less than
$\varepsilon $.

\begin{figure}
    \setcaptionmargin{40mm}
    %\onelinecaptionsfalse % if the caption is multiline
    \onelinecaptionstrue  % if the caption is one-line
    \centering
    \includegraphics[scale=0.75]{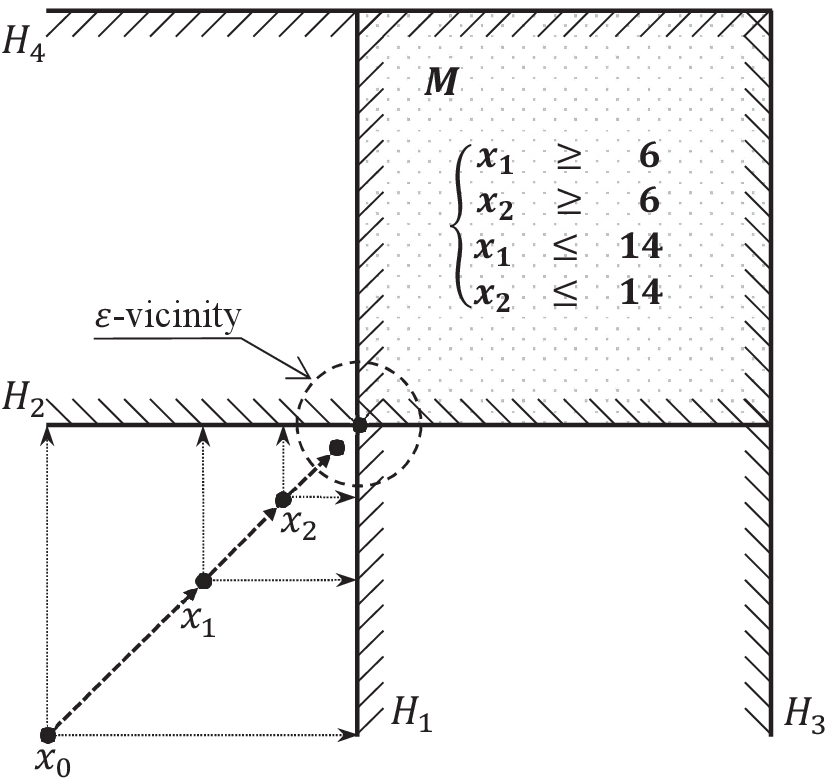}
    \captionstyle{normal}\caption{The case, when the rate of convergence slows down.}\label{Fig2}
\end{figure}

The pseudo-projection algorithm for the non-stationary case is shown in Fig.~\ref{Fig1}. Here $\varepsilon $ is a small positive value that is a parameter of the algorithm. This algorithm calculates a point $\tilde x$ that is an approximate solution of the non-stationary system of linear inequalities~\eqref{Formula1} in the sense that the point $\tilde x$ belongs to the polytope ${M^{(t)}}$ with precision $\varepsilon $ (Step~4). If the polytope $M$ does not change over time, the Algorithm~1 finishes in a finite number of steps for any $\varepsilon  > 0.$ This follows from the fact that the mapping ${\alpha ^{(t)}} = {x_k} - {\varphi ^{(t)}}\left( {{x_k}} \right)$ used in the Step~3, when $t$ is fixed, is an single-valued continuous ${M^{(t)}}$-fejerian mapping~\cite{13}, and consequently the iterative process implemented in the Algorithm~1 converges to a point belonging to the polytope ${M^{(t)}}$~\cite{14}. Another proof of the convergence of the Algorithm~1 can be found in~\cite{3}. However, both proofs are valid only for stationary problems. The non-stationary case was considered by us in~\cite{2} under the assumption that changing the initial data is a translation of the polytope~${M^{(t)}}.$ For this case, we have proved the theorem of a sufficient condition of converging the iterative process implemented by the Algorithm~1.

We have performed a parallel implementation of the Algorithm~1 using the BSF-skeleton~\cite{15} and have executed large-scale computational experiments on the ``Tornado SUSU''~\cite{16} computing cluster. We simulated the non-stationarity of the original inequality system by displacing the polytope ${M^{(t)}}$ along a straight line at a given rate. The obtained results show that Algorithm~1 demonstrates good scalability of up to 200 processor nodes when solving a non-stationary system involving 108002 inequalities and 54000 variables. However, in some cases, the maximum rate of the polytope displacement, at which the algorithm converged, did not exceed two units per second. It is insufficient for solving practical problems. Such result can be explained by the simple example shown in Fig.~\ref{Fig2}. A peculiarity of the Fejer mapping ${\alpha ^{(t)}} = {x_k} - {\varphi ^{(t)}}\left( {{x_k}} \right)$ used in Step~3 is that each new approximation is closer to the polytope than the previous one. But, in the case shown in Fig.~\ref{Fig2}, the Fejer process never reaches the polygon $M$. The most we can achieve in a finite number of iterations is to get a point inside of the polygon vertex vicinity. Assume, that at time ${t_1}$, the Algorithm~1 starts to perform the iterative process (Step~3) giving an approximate solution $\tilde x$ with distance to the polytope ${M^{({t_1})}}$ less than $\varepsilon $ (we used $\varepsilon  = {10^{ - 7}}$ in the experiments). Let the calculation be stopped at time ${t_2}$. During the elapsed time, the polytope is displaced to the position ${M^{({t_2})}}$, the distance from which to the point $\tilde x$ is more than $\varepsilon $. Therefore, we have to start the calculation again. In such a way, we will never get a solution with given precision.

\begin{figure}
    \setcaptionmargin{40mm}
    %\onelinecaptionsfalse % if the caption is multiline
    \onelinecaptionstrue  % if the caption is one-line
    \centering
    \includegraphics[scale=0.8]{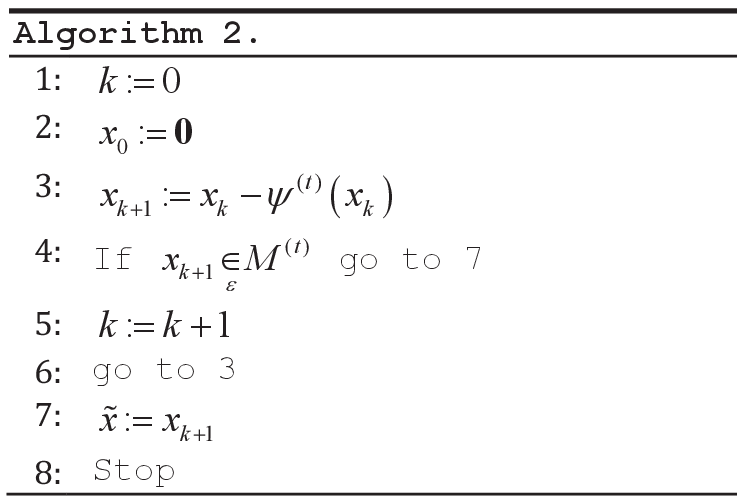}
    \captionstyle{normal}\caption{Modified pseudo-projection algorithm (ModAP).}\label{Fig3}
\end{figure}

In order to repair the problem, we replaced the mapping ${\varphi ^{(t)}}$ used in Step~3 of Algorithm~1 with the following mapping ${\psi ^{(t)}}$:
\begin{equation}\label{Formula12}{\psi ^{(t)}}\left( x \right) = \lambda \frac{{{\varphi ^{(t)}}\left( x \right)}}{{\left\| {{\varphi ^{(t)}}\left( x \right)} \right\|}},\end{equation}
where $\lambda $ is a positive constant being a parameter of the algorithm. A peculiarity of the mapping ${\psi ^{(t)}}$ is that its result is always a vector of fixed length $\lambda $. The modified algorithm, named ModAP, is presented in Fig.~\ref{Fig3}.

\section{Parallel version of ModAP algorithm} \label{section3}

To construct a parallel version of the algorithm ModAP, we use the BSF (Bulk Synchronous Farm) model of parallel computations~\cite{17}. In accordance to the technique proposed in~\cite{18}, we represent the algorithm in the form of operations on lists using the higher-order functions $Map$ and $Reduce$ defined in the Bird-Meertens formalism~\cite{19}. Let us define the list $\mathbb{A} = \left[ {({a_1},{b_1}), \ldots ,({a_m},{b_m})} \right]$, where ${a_i}$ is the $i$-th row of the matrix $A$, and ${b_i}$ is the $i$-th element of the column $b$ $(i = 1, \ldots ,m)$. Here, we omit the upper index of time $(t)$ assuming the moment of time to be fixed.

\begin{figure}
    \setcaptionmargin{40mm}
    %\onelinecaptionsfalse % if the caption is multiline
    \onelinecaptionstrue  % if the caption is one-line
    \centering
    \includegraphics[scale=0.8]{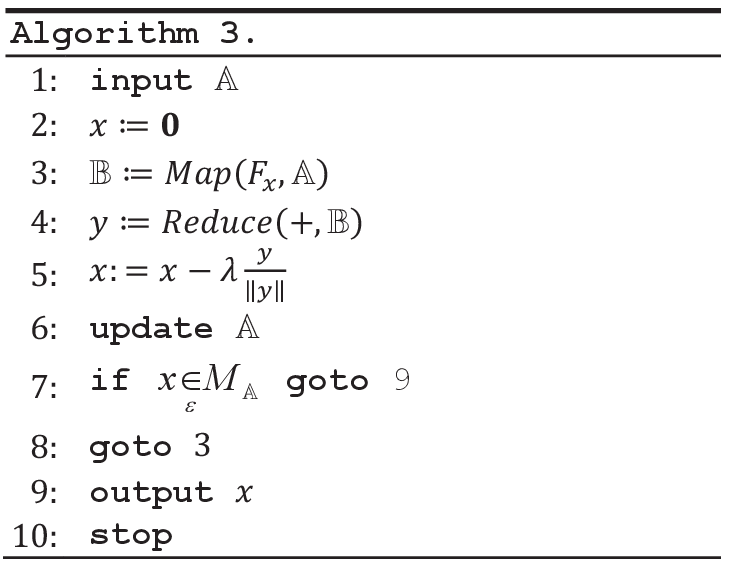}
    \captionstyle{normal}\caption{ModAP algorithm on lists.}\label{Fig4}
\end{figure}

Let us define the parameterized function ${F_x}:{\mathbb{R}^n} \times \mathbb{R} \to {\mathbb{R}^n} \times \{ 0,1\} $ as follows:
\begin{equation}\label{Formula13}{F_x}\left( {{a_i},{b_i}} \right) = \rho _{{H_i}}^ + \left( x \right).\end{equation}
This function maps the pair $\left( {{a_i},{b_i}} \right)$ to the
vector  ${y_i} = \rho _{{H_i}}^ + (x)$, which is the positive slice
of the reflection vector of the point $x$ relative to the hyperplane
${H_i} = \left\{ {x \in {\mathbb{R}^n}|\left\langle {{a_i},x}
\right\rangle  = {b_i}} \right\}$. The higher-order function
$Map({F_x},\mathbb{A})$ applies the function ${F_x}$ to each element
of list $\mathbb{A}$ converting it to the following list
\begin{equation*}\label{Formula14}
\mathbb{B} = \left[ {{y_1}, \ldots ,{y_m}} \right] = \left[ {{F_x}({a_1},{b_1}), \ldots ,{F_x}({a_m},{b_m})} \right],
\end{equation*}
where ${y_i} \in {\mathbb{R}^n}$ $(i = 1, \ldots ,m)$.  Let the
symbol $ + $ denote vector addition operation. Then, the
higher-order function $Reduce( + ,\mathbb{B})$ calculates the vector
y, which is the sum of the vectors of the list~$\mathbb{B}$:
%\begin{equation}\label{Formula15}
$y = {y_1} +  \ldots  + {y_m}.$
%\end{equation}
Obviously, according
to the equation~\eqref{Formula8}, we have
%\begin{equation}\label{Formula16}
$\varphi \left( x \right) = y/h.$
%\end{equation}
Taking~\eqref{Formula12} into account, we obtain from this
\begin{equation*}\label{Formula17}
\psi \left( x \right) = \lambda \frac{y}{{\left\| y \right\|}}.
\end{equation*}
In such a way, we obtain the sequential version of the ModAP
algorithm on lists presented in Fig.~\ref{Fig4}. In Step~1, the
input of the list $\mathbb{A}$ containing initial values of the
matrix $A$ and the column $b$ of the inequality
system~\eqref{Formula1} are executed. In Step~2, the zero vector is
assigned to the $x$ as an initial approximation. In Step~3, the
$Map$ function calculates the list~$\mathbb{B}$ of positive slices
of the reflection vectors of the point $x$ relative to the
hyperplanes bounding the polytope ${M_\mathbb{A}}$, which is the
feasible region of the inequality system~\eqref{Formula1}. In
Step~4, the vector $y$ is calculated as the sum of all positive
slices of the reflection vectors. Step~5 calculates the next
approximation $x$. In Step~6, the original data of the inequality
system~\eqref{Formula1} is updated. Step~7 checks if the obtained
approximation belongs to the polytope ${M_\mathbb{A}}$ with
precision $\varepsilon $. If so, then we go to Step~9, where $x$ is
output as the approximate solution, after which the iterative
process is stopped. Otherwise, we go from Step~8 to Step~3, and the
iterative process continues.

The parallel version of the modified pseudo-projection algorithm on
lists (ModAPL) is presented in Fig.~\ref{Fig5}.  We used the
master-slave paradigm~\cite{20}. It is assumed that the computing
system includes one master and $K$ slaves ($K \geq 1$). In Step~1,
the master and all slaves load the source data presented as the list
$\mathbb{A}$. Step~3 starts the iterative process: the master sends
the current approximation $x$ to all slaves. In Step~4, the slaves
independently calculate their part of the list~$\mathbb{B}$. In
Step~5, the slaves independently sum up the elements of their
sublists of the list~$\mathbb{B}$. In Step~6, the calculated vectors
${y^{(1)}}, \ldots ,{y^{(K)}}$ are sent to the master. In Step~7,
the master calculates the resulting vector $y$. Step~8 calculates
the next approximation $x$. In Step~9, the original data of the
inequality system~\eqref{Formula1} is updated by the master and all
slaves. Step~10 checks the stopping criterion and assigns the
appropriate value to the variable exit. In Step~11, the master sends
the value of the variable exit to all slaves. Depending on this
value, we either go to the next iteration or terminate the iterative
process (Steps 12--16).

\section{Analytical study of parallel algorithm} \label{section4}

\begin{figure}
    \setcaptionmargin{40mm}
    %\onelinecaptionsfalse % if the caption is multiline
    \onelinecaptionstrue  % if the caption is one-line
    \centering
    \includegraphics[scale=0.8]{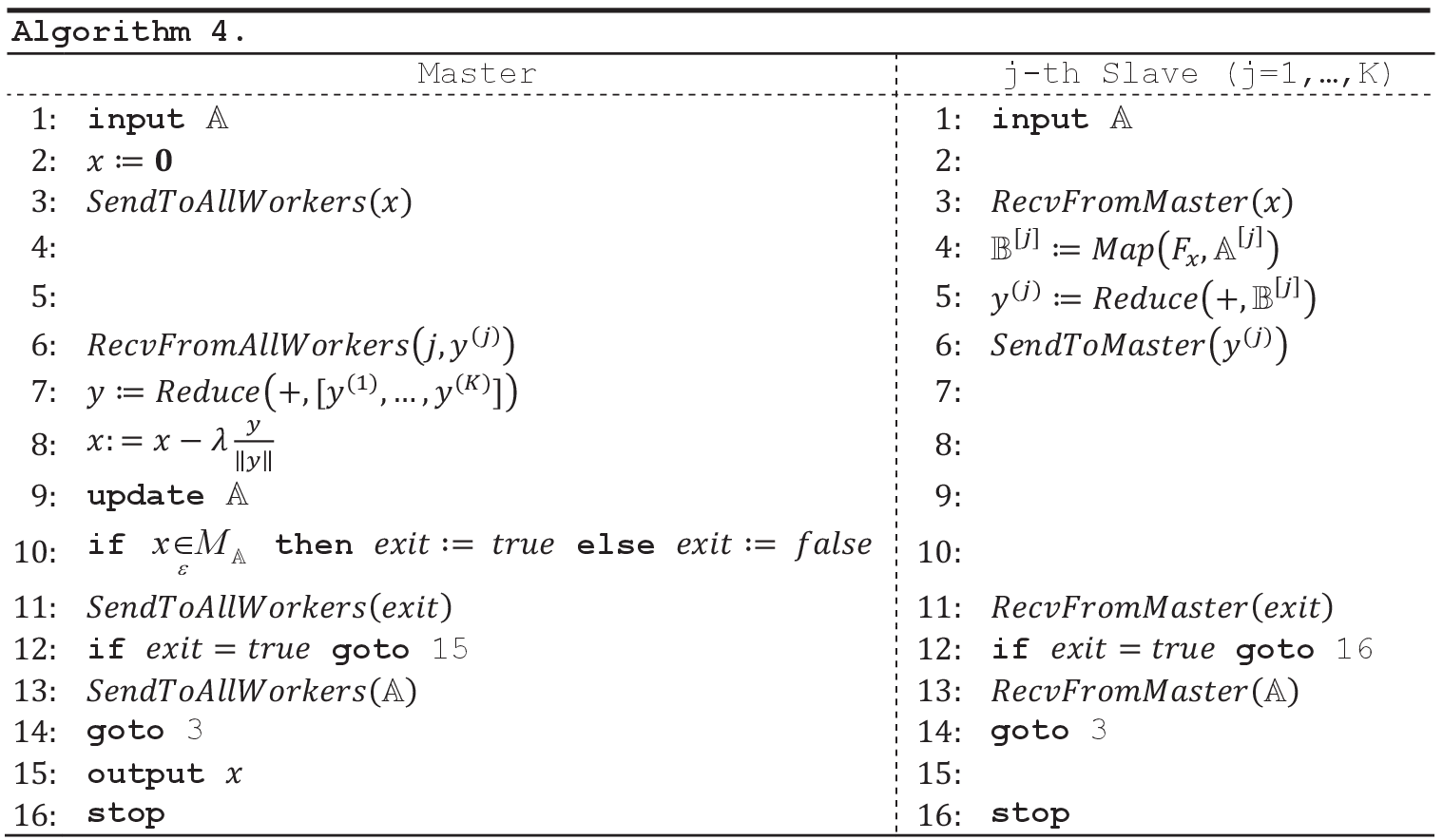}
    \captionstyle{normal}\caption{ModAPL parallel algorithm on lists.}\label{Fig5}
\end{figure}

To obtain an analytical estimation of the scalability of the
parallel version of  the ModAPL algorithm (Fig.~\ref{Fig5}) we use
the cost metric of the BSF model~\cite{18}. It includes the
following parameters:
\begin{tabbing}
MM. \= M \= MMMMMMMMMMMMMMMMMMMMMMMMMMMMMMMMMMMMMMMMMMMMMMMMMMMMMMMMMMMMMMM \kill
$K$ \> : \> number of slave nodes;\\
${t_s}$ \> : \> time spent by the master node to send current approximation to one slave node\\
\>\>(excluding latency);\\
${t_{Map}}$ \> : \> time spent by a single slave node to process the entire list $\mathbb{A}$;\\
${t_p}$ \> : \> time spent by the master node to process results received from the slave nodes and to\\
\>\>check the stopping criterion;\\
${t_r}$ \> : \> time spent by the master node to receive the result from one slave node (excluding\\
\>\>latency);\\
${t_a}$ \> : \> time spent by a node (master or slave) to process an addition of two vectors;\\
$l$ \> : \> length of the list $\mathbb{A}$ (the same as the length of the list $\mathbb{B}$);\\
$L$ \> : \> latency (time of transferring one-byte message node-to-node).\\
\end{tabbing}

For the ModAPL algorithm we have
\begin{equation}\label{Formula18}l = m.\end{equation}

Within one iteration, we introduce the following notation:
\begin{tabbing}
MM. \= M \= MMMMMMMMMMMMMMMMMMMMMMMMMMMMMMMMMMMMMMMMMMMMMMMMMMMMMMMMMMMMMMM \kill
$c_s$\> : \> number of float values sending from the master node to one slave node;\\
$c_{Map}$\> : \> number of arithmetic operations to execute the higher-order function $Map$ for the\\
\>\>entire list $\mathbb{A}$;\\
${c_a}$\> : \> number of arithmetic operations to add two vectors in $n$-dimensional space;\\
$c_r$\> : \> number of float values sending from one slave node to the master node;\\
${c_p}$\> : \> number of arithmetic operations performed by the master in Steps 8 and 10 of\\
\>\>Algorithm~4;\\
${c_u}$\> : \> number of float values sending from the master node to one slave node in Step~13 of\\
\>\>Algorithm~4.
\end{tabbing}

Calculate the specified values. At the beginning of each iteration
in Step~3, the  master sends the vector $x$ containing $n$ float
values to every slave. Therefore
%\begin{equation}\label{Formula19}
${c_s} = n.$
%\end{equation}
 In the $Map$ step (Step~3 of the
Algorithm~3), the function ${F_x}$ defined by the
equation~\eqref{Formula13} is applied to all elements of the list
$\mathbb{A}$ having the length $m$. In view of~\eqref{Formula7},
this implies
%\begin{equation}\label{Formula20}
${c_{Map}} = (5n + 1)m.$
%\end{equation}
To add two vectors of
dimension $n$, one must execute $n$ arithmetic operations.
Therefore,
%\begin{equation}\label{Formula21}
${c_a} = n.$
%\end{equation}
In Step~6 of Algorithm~4, $j$-th slave
node sends the vector ${y^{(j)}}$ of dimension $n$ to the master
node. Hence it follows that
%\begin{equation}\label{Formula22}
${c_r} = n.$
%\end{equation}
Assume that the square root is calculated
by using the first four terms of the Taylor series. This is 9
arithmetic operations. Then, the execution of Step~8 of Algorithm~4
requires $5n + 8$ arithmetic operations. According to the
equation~\eqref{Formula10}, the execution of Step~11 requires $(6n +
11)m$ arithmetic operations. Hence it follows that
\begin{equation*}\label{Formula23}
{c_p} = (6n + 11)m + 5n + 8.
\end{equation*}
If only one value changes in the source data during one iteration, then one real number must be sent in Step~13. In this case, we have
%\begin{equation}\label{Formula24}
${c_u} = 1.$
%\end{equation}
If all values changes in the source data during one iteration, then $(n + 1)m$ real number must be sent in Step~13. In this case, we have
%\begin{equation}\label{Formula25}
${c_u} = (n + 1)m.$
%\end{equation}

Let ${\tau _{op}}$ denote the time that a processor node spends to execute one arithmetic operation (or one comparison operations), and ${\tau _{tr}}$ denote the time that a processor node spends to send one float value to another processor node  excluding latency. Then for ${c_u} = 1$ we get the following values for the cost parameters of the ModAPL parallel algorithm:
\begin{equation}\label{Formula26}{t_s} = ({c_s} + {c_u}){\tau _{tr}} = (n + 1){\tau _{tr}};\end{equation}
\begin{equation}\label{Formula27}{t_{Map}} = {c_{Map}}{\tau _{op}} = (5n + 1)m{\tau _{op}};\end{equation}
\begin{equation}\label{Formula28}{t_r} = {c_r}{\tau _{tr}} = n{\tau _{tr}};\end{equation}
\begin{equation}\label{Formula29}{t_a} = {c_a}{\tau _{op}} = n{\tau _{op}};\end{equation}
\begin{equation}\label{Formula30}{t_p} = {c_p}{\tau _{op}} = ((6n + 11)m + 5n + 8){\tau _{op}}.\end{equation}
Using the scalability boundary equation  from~\cite{18}, and taking
into account~\eqref{Formula18}, we obtain
from~\eqref{Formula26}--\hspace{0pt}\eqref{Formula30} the following
estimation:
\begin{equation}\label{Formula31}{K_{MAX}} = \sqrt {\frac{{{t_{Map}} + l{t_a}}}{{2L + {t_s} + {t_r} + {t_a}}}}  = \sqrt {\frac{{(5n + 1)m{\tau _{op}} + nm{\tau _{op}}}}{{2L + (n + 1){\tau _{tr}} + n{\tau _{tr}} + n{\tau _{op}}}}} .\end{equation}
Assuming $m = kn$ for some $k \in \mathbb{N}$ and $n \gg 1$, it is easy to get the following estimation from~\eqref{Formula31}:
\begin{equation}\label{Formula32}{K_{MAX}} = O\left( {\sqrt n } \right),\end{equation}
where ${K_{MAX}}$ is the number of processor nodes, for which the maximal speedup is achieved.

When ${c_u} = (n + 1)m$ we have ${t_s} = ({c_s} + {c_u}){\tau _{tr}} = (n + (n + 1)m){\tau _{tr}}$. We obtain from this the following estimation:
\begin{equation}\label{Formula33}{K_{MAX}} = \sqrt {\frac{{{t_{Map}} + l{t_a}}}{{2L + {t_s} + {t_r} + {t_a}}}}  = \sqrt {\frac{{(5n + 1)m{\tau _{op}} + nm{\tau _{op}}}}{{2L + (n + (n + 1)m){\tau _{tr}} + n{\tau _{tr}} + n{\tau _{op}}}}} .\end{equation}
Assuming $m = kn$ for some $k \in \mathbb{N}$ and $n \gg 1$, it is easy to get the following estimation from~\eqref{Formula33}:
\begin{equation}\label{Formula34}{K_{MAX}} = O\left( 1 \right).\end{equation}
Thus, we can conclude that the ModAPL parallel algorithm will demonstrate a limited scalability when a small part of the source data is dynamically changed. In this case, according to the equation~\eqref{Formula32}, the scalability bound increases proportionally to the square root of $n$, where $n$ is the space dimension. If a large part of the source data is dynamically changed, then, according to the equation~\eqref{Formula34}, the ModAPL parallel algorithm becomes inefficient due to the lack of speedup.

\section{Computational experiments} \label{section5}

\begin{figure}
    \setcaptionmargin{40mm}
    %\onelinecaptionsfalse % if the caption is multiline
    \onelinecaptionstrue  % if the caption is one-line
    \centering
    \includegraphics[scale=1]{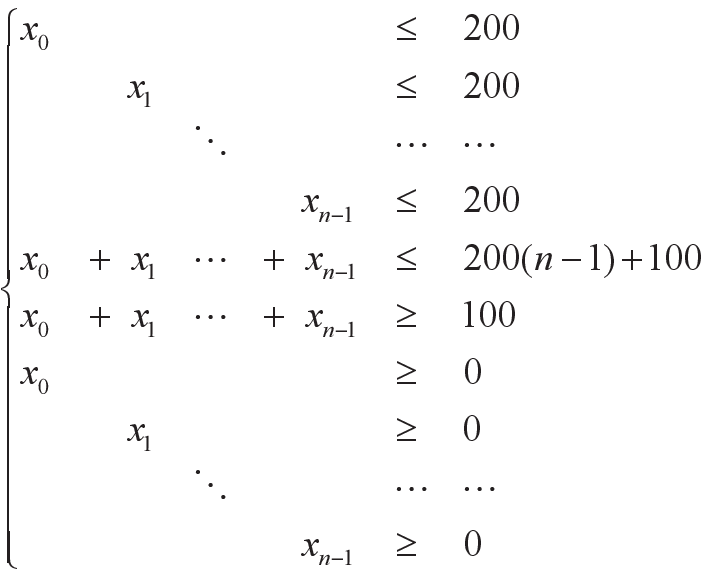}
    \captionstyle{normal}\caption{ Test problem.}\label{Fig6}
\end{figure}

\begin{figure}
    \setcaptionmargin{32mm}
    \onelinecaptionsfalse % if the caption is multiline
    %\onelinecaptionstrue  % if the caption is one-line
    \centering
    \includegraphics[scale=0.8]{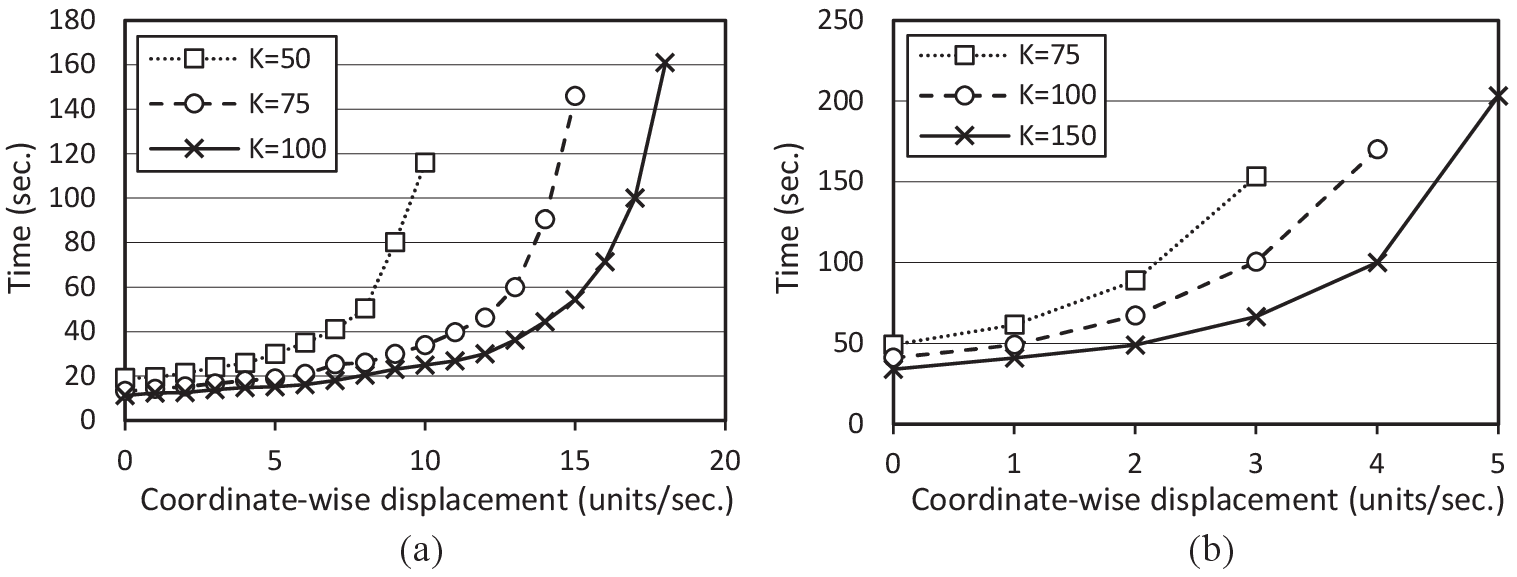}
    \captionstyle{normal}\caption{Experiment results ($K$ is the number of processor nodes):\\
(a) $n = 32000$ and $m = 64002$;
(b) $n = 54000$ and $m = 108002$.}\label{Fig7}
\end{figure}

To verify the efficiency of the ModAPL parallel algorithm, we
implemented this algorithm in C++ language  using a parallel
BSF-skeleton~\cite{15} based on MPI parallel programming library.
All source codes are freely available at
\url{https://github.com/leonid-sokolinsky/NSLP-Quest}. As a test
problem, we used a scalable inequality system of dimension $n$
from~\cite{21}. This system is presented in Fig.~\ref{Fig6}. The
number of inequalities in this system is $m = 2n + 2$. The
non-stationarity of the problem was simulated by a translation of
the polytope ${M^{(t)}}$. The computational experiments were
conducted on the ``Tornado SUSU'' computing cluster~\cite{16}, whose specifications are shown in Table~\ref{Table1}. The
results of the experiments are presented in Fig.~\ref{Fig7}. The diagrams show the
dependence of the running time of the ModAPL parallel algorithm on
the displacement rate of the polytope~${M^{(t)}}$. The values of
coordinate-wise displacement indicate how many units per second will
be added to each coordinate of the polytope $M$ before next
iteration. In the first experiment, we used the system in
Fig.~\ref{Fig6} with dimension $n = 32000$ and number of
inequalities $m = 64002$. Calculations were conducted on the
hardware configurations with 50, 75 and 100 processor nodes. The
results of this experiment are presented in Fig.~\ref{Fig7}~(a). In
the diagram, $K$ denotes the number of processor nodes in the
hardware configuration. For $K = 50$, the highest rate of the
coordinate-wise shift at which the iterative process ``catch up with
the polytope'' was approximately 10 units per second. Adding up the
displacement vectors for all coordinates, we get the total rate
vector with a length equal to $\sqrt {32000 \cdot {{10}^2}}  \approx
{\text{1789}}$ units per second. For $K = 75$, the highest total
displacement rate at which the algorithm converged increases to 2683
units per second. And for $K = 100$, the convergence of the
algorithm was observed at the total displacement rate of up to 3220
units per second. For $K > 100$, the parallel efficiency for the
problem of dimension $n = 32000$ began to fall off.

\begin{table}[!htb]
\setcaptionmargin{0mm}
\onelinecaptionstrue
\captionstyle{flushleft}
\caption{Specifications of ``Tornado SUSU'' computing cluster.}
\bigskip
\begin{tabular}{l|l}
  \hline
  Number of processor nodes & 480 \\
  Processor & Intel Xeon X5680 (6 cores 3.33 GHz) \\
  Processors per node & 2\\
  Memory per node & 24 GB DDR3\\
  Interconnect & InfiniBand QDR (40 Gbit/s) \\
  Operating system & Linux CentOS\\[1mm]
  \hline
\end{tabular}\label{Table1}
\end{table}

In the second experiment, we used the system in Fig.~\ref{Fig6} with
dimension $n = 54000$ and  number of inequalities $m = 108002$.
Calculations were conducted on the hardware configurations with 75,
100 and 150 processor nodes. The results of this experiment are
presented in Fig.~\ref{Fig7}~(b). In this case, the maximum
displacement rates of the polytope at which the algorithm converged
were 697, 930, and 1162 units per second, respectively. For $K >
150$, the parallel efficiency for the problem of dimension $n =
54000$ also began to fall off.

Thus, the conducted experiments show that the proposed modified algorithm of pseudo-projections allows the effective parallelization on cluster computing systems and is able to find feasible points for non-stationary systems of linear inequalities that dynamically changed in a certain way with high rate.

\section{Conclusion} \label{section6}

In this article, we presented the modified parallel iterative
pseudo-projection algorithm that can find feasible points for
non-stationary systems of linear inequalities of large dimensions on
cluster computing systems. This algorithm was presented as the
ModAPL algorithm on lists using the higher-order functions $Map$ and
$Reduce$. For the ModAPL algorithm, we obtained an estimation of the
scalability bound of its parallel version using the cost metric of
the BSF parallel computing model~\cite{18}. If a small part of the
source data is dynamically changed during one iteration then the
scalability bound is equal to $O\left( {\sqrt n } \right)$, where
$n$ is the problem dimension. If a large part of the source data is
dynamically changed during one iteration then the scalability bound
is equal to $O\left( 1 \right)$, which means there is no
acceleration at all. We implemented ModAPL parallel algorithm in C++
language using a parallel BSF-skeleton~\cite{15}. All source codes
are freely available at
\url{https://github.com/leonid-sokolinsky/NSLP-Quest}. By using this
implementation, we conducted the large-scale computational
experiments on a computing cluster. We simulated the problem
non-stationarity by transferring the polytope bounding the feasible
region by a certain number of units during each iteration. The
conducted experiments show that the ModAPL parallel algorithm is
able to find feasible points for non-stationary systems of linear
inequalities with 54~000 variables and 108~002 inequalities on a
cluster computing system with 200 processor nodes.

\begin{acknowledgments}
This research was partially supported by the Russian Foundation for Basic Research (project No.~20-07-00092-a), by the Ministry of Science and Higher Education of the Russian Federation (gov. order FENU-2020-0022) and by the Government of the Russian Federation according to Act 211 (contract No.~02.A03.21.0011).
\end{acknowledgments}

% Text of article ends here.

%
% The Bibliography (Mendeley style IEEE)
%

\end{document}